\documentclass[prl,twocolumn,amssymb,showpacs]{revtex4}

\usepackage{epsfig}

\begin{document}

\title{Photodetachment of cold OH$^-$ in a multipole ion trap}

\author{S. Trippel, J. Mikosch, R. Berhane, R. Otto,
 M. Weidem{\"u}ller, and R. Wester}
\email[]{E-mail: roland.wester@physik.uni-freiburg.de}

\affiliation{Physikalisches Institut, Universit{\"a}t Freiburg,
  Hermann-Herder-Stra{\ss}e 3, 79104 Freiburg, Germany}

\date{\today}


\begin{abstract}
\noindent
The absolute photodetachment cross section of OH$^-$ anions at a rotational
and translational temperature of 170\,K is determined by measuring the
detachment-induced decay rate of the anions in a multipole radio-frequency ion
trap. In comparison with previous results, the obtained cross section shows
the importance of the initial rotational state distribution. Using a
tomography scan of the photodetachment laser through the trapped ion cloud,
the derived cross section is model-independent and thus features a small
systematic uncertainty. The tomography also yields the column density of the
OH$^-$ anions in the 22-pole ion trap in good agreement with the expected
trapping potential of a large field free region bound by steep potential
walls.
\end{abstract}




\maketitle


Photodetachment of the excess electron from a negative ion represents a
fundamental light-matter interaction process that reveals detailed information
on atomic and molecular structure and dynamics. The additional electron is
only bound by virtue of electron-electron interactions and its binding energy
often depends sensitively on electron correlations in the entangled
multi-electron wavefunction. Spectroscopic information, obtained from
photoelectron energy measurements, is used to study the electronic,
vibrational and rotational Eigenstate spectrum of anionic and neutral
molecules. This also yields insight in transition states structures of
reactive collisions of neutral molecules \cite{neumark2005:pccp}. In addition,
ultrafast time-resolved studies of wavepacket or electron rearrangement
dynamics inside clusters have employed anion photoelectron spectroscopy
\cite{stolow2004:cr}. Multiphoton electron detachment in short laser pulses
provides information on the electron-atom interaction in strong laser fields
\cite{reichle2001:prl}.  Cross sections for photodetachment reveal
complementary information to photoelectron spectra and serve to challenge
calculations for bound-free transition matrix elements
\cite{soerensen1995:ps}.

The most detailed photodetachment cross section studies on a molecular anion
have been carried out for the hydroxyl anion OH$^-$. Much of the work on
OH$^-$ has focused on relative cross sections. Near threshold, comparison of
the relative cross section as a function of electron energy with Wigner
threshold laws allows for precise tests of the long-range electron-neutral
interactions. Specifically, the subtle coupling of the two lowest
$\Lambda$-doublet states has been observed, which leads to a long-range
electron-dipole interaction \cite{schulz1982:jcp,smith1997:pra}. Also,
relative cross section measurements for transitions of specific rotational
states have been carried out for OH$^-$
\cite{delsart2002:prl,goldfarb2004:jcp}. These results indicate deviations of
the photodetachment cross section from s-wave electron detachment
calculations. Interpretation, however, is obscured by the incomplete knowledge
of the OH$^-$ rotational population. Even for the OH$^-$ anion only two
absolute cross section measurements have been performed to date
\cite{branscomb1966:pra,lee1979:jcp}, which disagree with each other and
deviate from the calculated one by almost an order of magnitude
\cite{soerensen1995:ps}.

An important application of rotational-state dependent absolute cross sections
are models of the negative ion abundance in planetary atmospheres as well as
in interstellar molecular clouds, because in these low-temperature
environments photodetachment by cosmic radiation represents an important loss
channel for negative ions. In the interstellar medium, negative ions have not
been detected up to now \cite{morisawa2005:paj}, but their importance is still
under debate \cite{petrie1997:apj}.

In this letter we present a scheme to measure absolute photodetachment cross
sections of trapped molecular anions. The combination of negative ion
photodetachment with cooling and trapping of molecular anions in a
radio-frequency multipole trap \cite{gerlich1995:ps} allows us to study
photodetachment with a well-controlled rotational state distribution.  The
scheme yields absolute cross sections with little systematic uncertainties and
without resorting to theoretical models. We apply the method to OH$^-$ anions
that are trapped in a 22-pole rf ion trap and cooled in a helium buffer gas to
a temperature of 170\,K. We show that the obtained absolute cross section
value only agrees, within the experimental accuracy, with previous knowledge,
if its dependence on the OH$^-$ rotational state is taken into account, which
exemplifies the importance of controlling the rotational temperature.

A number of previous photodetachment studies have employed negative ions in
traps. In a Penning trap, the photodetachment of H$^-$ near threshold
\cite{harms1997:jpb} and of multiply charged metal cluster anions has been
observed \cite{herlert2000:hi}. The lifetime of dipole-bound negative ions has
been found to be limited by blackbody radiation induced photodetachment
\cite{suess2003:jcp}. Photodetachment of gold anions in a hyperbolic Paul trap
revealed the existence of resonances in the cross section
\cite{champeau1998:jpb1}. Metastable states of negative ions have been studied
with photodetachment spectroscopy as a function of storage time in a Zajfman
ion trap \cite{naaman2000:jcp}. The high-order multipole radio-frequency trap
that is employed in this work allows trapping of molecular ions in an
approximate box potential \cite{gerlich1995:ps}, in contrast to the harmonic
confinement of the classical Paul trap that is used for a precise localization
and manipulation of single ions \cite{leibfried2003:rmp}. Multipole ion traps
have attracted attention, because in buffer gas cooling of molecules and
clusters thermal equilibrium is reached not only for the translational, but
also for the internal degrees of freedom for temperatures down to 10\,K. These
traps are therefore applied in laboratory astrophysics \cite{gerlich2006:ps},
highly sensitive electronic \cite{boyarkin2006:jacs} and infrared
\cite{asvany2005:sci,mikosch2004:jcp} spectroscopy, and as a source of cold
molecular ions for collision studies \cite{kreckel2005:prl}.

Photodetachment of an ensemble of $N$ trapped negative ions leads to an
additional loss channel for the trapped ions. The rate equation for the number
of trapped ions is obtained by integrating in cylindrical coordinates over the
rate equation for the spatial density $n$ of the ions in the trap,
\begin{equation}
\frac{dN}{dt} = - \int \sigma_{\rm pd} \Phi(r,\phi) n(r,z) r dr d\phi dz -
\Gamma N
\label{rate_equation1:eq}
\end{equation}
where $\sigma_{\rm pd}$ is the total cross section for OH$^-$ photodetachment
at the employed laser wavelength. $\Phi$ is photon flux density of the laser,
which is taken to be constant along the laser beam parallel to the
$z$-axis. $\Gamma$ denotes the loss rate due to residual gas
collisions. Cylindrical symmetry is assumed for the trapping potential so that
the ion density $n$ becomes independent of the angle $\phi$.  Assuming further
a dilute ion cloud with negligible ion-ion interaction, the integration of the
density $n(r,z)$ over the $z$-direction yields the particle number $N$ times a
single-particle column density $\rho(r)$. The remaining two-dimensional
overlap integral in Eq.\ \ref{rate_equation1:eq} is evaluated by assuming a
narrowly focused laser beam such that the ion density $\rho(r_{\rm L})$ at the
position of the laser beam is constant over the laser beam cross section and
can thus be moved in front of the integral. The integral is then given by the
total photon flux through the ion trap $F_{\rm L} = \int \Phi(r,\phi) r dr
d\phi$, and one obtains
\begin{equation}
\frac{dN}{dt} = 
 - [ \sigma_{\rm pd} \rho(r_{\rm L}) F_{\rm L} + \Gamma ] N.
\label{rate_equation3:eq}
\end{equation}
This equation is solved by an exponential decay with a rate $k_{\rm pd}(r_{\rm
L}) + \Gamma$ given by the square brackets, which is the quantity that can be
measured experimentally.

\begin{figure}[tb]
  \begin{center}
    \includegraphics[width=0.9 \columnwidth]{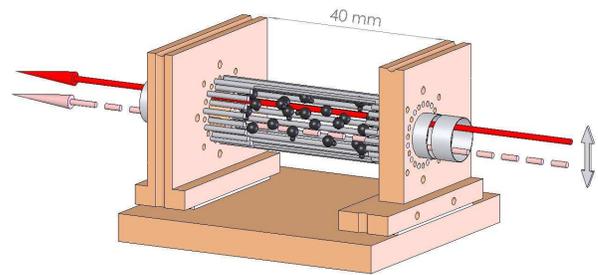}
    \caption{Schematic view of the 22-pole ion trap with the rf electrodes for
      radial and the cylindrical dc end-cap electrodes for axial
      confinement. The position of the photodetachment laser is scanned along
      the vertical axis.}
    \label{trap:fig}
  \end{center}
\end{figure}

To back out the absolute cross section, the ion column density $\rho(r_{\rm
L})$ needs to be determined. This is achieved by scanning the photodetachment
laser through the distribution of trapped ions and measuring at each laser
position $r_{\rm L}$ the decay rate $k_{\rm pd}(r_{\rm L})$, which is
proportional to $\rho(r_{\rm L})$ according to Eq.\
\ref{rate_equation3:eq}. With the normalization condition for $\rho(r_{\rm
L})$ one obtains directly the absolute photodetachment cross section
\begin{equation}
\sigma_{\rm pd} = \frac{1}{F_{\rm L}} \int k_{\rm pd}(r) r dr d\phi
\label{cross_section:eq}
\end{equation}
and the absolute single-particle column density to find an ion at a radius
$r_{\rm L}$
\begin{equation}
\rho(r_{\rm L}) = \frac{k_{\rm pd}(r_{\rm L})}
                           {\int k_{\rm pd}(r) r dr d\phi}.
\label{ion_density:eq}
\end{equation}
Thus the determination of both the cross section and the density only requires
measurements of quantities that are directly accessible in an experiment.


In the experiment, OH$^-$ anions are stored in a 22-pole ion trap
\cite{gerlich1995:ps}, shown schematically in Fig.\ \ref{trap:fig}. The trap
is cooled to 170\,K, which strongly increases the lifetime of the trapped
ions. A focused helium neon laser at 632.8\,nm is passed through the trap,
which photodetaches OH$^-$ (electron affinity 1.83\,eV
\cite{goldfarb2004:jcp}) in a one-photon transition. In the trap ions are
stored in a cylindrical structure of 22 stainless steel rods of 40\,mm length
and 10\,mm inscribed diameter, alternatingly connected to the two phases of a
radio-frequency oscillator. An effective potential is formed with a radial
scaling proportional to $r^{20}$, which leads to steep potential walls around
an almost field free trapping volume. In the axial direction the ions are
trapped by cylindrical entrance and exit electrodes of 7.1\,mm inner diameter.

The trap is loaded with a packet of about 10$^3$ NH$_2^-$ anions which are
chemically converted into OH$^-$ ions in the trap within a few seconds of
storage time. NH$_2^-$ ions are produced by dissociative attachment of slow
secondary electrons to NH$_3$ in a pulsed supersonic expansion of 90\% Ne and
10\% NH$_3$ that is ionized by a counter-propagating 2\,keV electron beam. The
NH$_2^-$ anions are accelerated, mass selected in a Wiley-McLaren
time-of-flight mass spectrometer \cite{wimc}, decelerated again, and focused
into the ion trap. Inside the trap they are collisionally cooled to energies
below the trapping potential by a high density pulse of helium buffer gas. A
continuous flow of helium buffer gas thermalizes the translational and
internal degrees of freedom of the trapped ions within a fraction of a second
with the temperature of the surrounding wall, which is cooled to 170\,K by
coupling to a liquid nitrogen bath. OH$^-$ ions are produced in the trap via
the reaction $\mathrm{NH}_2^- + \mathrm{H}_2\mathrm{O} \rightarrow
\mathrm{OH}^- + \mathrm{NH}_3$ by mixing traces of water to the helium buffer
gas \cite{smith1997:pra}. After an arbitrary time of storage the ions are
extracted from the trap by pulsing the trap exit electrode to an attractive
potential. The ions are then deflected towards a multichannel plate detector
which records time-of-flight spectra. Careful analysis of these spectra showed
that after 10\,s all NH$_2^-$ ions are converted into OH$^-$.

The photodetachment HeNe laser (Carl Zeiss Jena), which travels parallel to
the symmetry axis of the trap, has a wavelength of 632.8\,nm and a power of
$(3.0\pm0.1)$\,mW, which corresponds to a photon flux of $F_{\rm L} =
(9.5\pm0.3)\times 10^{15}$\,/s. With a telescope and a convex lens the beam is
focused to a waist of $210\,\mathrm{\mu m}$ which is placed in the center of
the trap. The laser beam is moved radially through the trap by moving the
convex lens with a translation stage. The laser light is switched on and off
via an electro-mechanic shutter to perform consecutive measurements with and
without laser interaction.

\begin{figure}[tb]
  \begin{center}
    \includegraphics[width=0.9 \columnwidth]{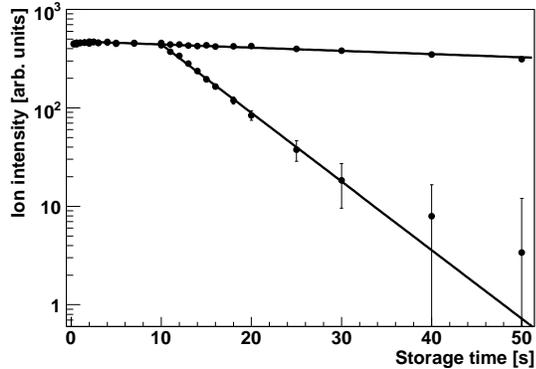}
    \caption{Measured OH$^-$ ion signal as a function of storage time in the
    trap. The upper set of points shows the signal in absence of the
    photodetachment laser. The lower points show that, when the laser is
    switched on at 10\,s, a fast additional loss channel due to
    photodetachment appears. The decay rates are obtained from exponential
    fits (solid lines).}
    \label{oh-photodet:fig}
  \end{center}
\end{figure}

Figure \ref{oh-photodet:fig} shows the measured ion intensity as a function of
storage time. The upper curve corresponds to a background measurement without
laser interaction. It yields a loss rate due to residual gas collisions of
$\Gamma = (7.5\pm0.3)\times 10^{-3}$\,s$^{-1}$ corresponding to a lifetime of
the trapped ions of $(133\pm6)$\,s. The lower curve shows the ion intensity as
a function of storage time when the laser, positioned along the symmetry axis
of the trap, is switched on 10\,s after loading the trap. The striking
difference in the loss rate from the trap is caused by photodetachment of the
negative ions by the laser photons, the only possible laser-driven process.

\begin{figure}[tb]
  \begin{center}
    \includegraphics[width=0.9 \columnwidth]{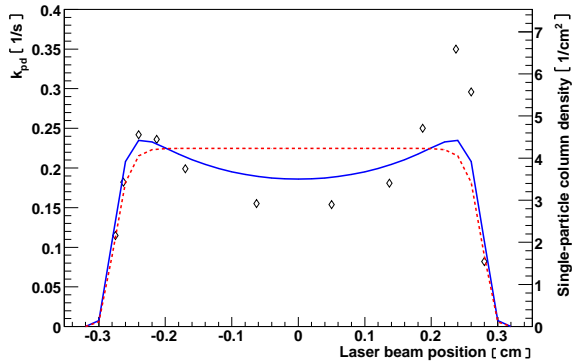}
    \caption{Measured photodetachment rate (left axis) and derived ion column
    density (right axis) as a function of the laser beam position (open
    symbols, their size represents the obtained accuracy). The solid (dashed)
    line shows the calculated column density taking into account (neglecting)
    the dependence of the trapping potential on the $z$-direction.}
    \label{particledistr:fig}
  \end{center}
\end{figure}

To determine the absolute cross section with Eq.\ \ref{cross_section:eq}, a
measurement of the laser-induced detachment rate $k_{\mathrm{pd}}(r)$ is
performed as a function of the radial position $r$ of the laser, as shown in
Fig.\ \ref{particledistr:fig} (left vertical axis). Using Eq.\
\ref{ion_density:eq} the detachment rate is normalized to reflect the
$r$-dependent ion column density (right vertical axis in Fig.\
\ref{particledistr:fig}). This measurement shows that the ion density drops
off near $\pm 3$\,mm. It is experimentally verified that this density
drop is not introduced by clipping of the laser beam at the end-cap electrodes
by expanding the trapped ion cloud in a weaker trapping potential.

To further analyze the shape of the measured density distribution of Fig.\
\ref{particledistr:fig}, a calculation of the ion density in the trap is
performed. Coulomb repulsion between the ions is expected to be too weak to
produce the maxima in the ion distribution, given that the plasma parameter is
estimated to be about 10$^{-4}$. We therefore calculate the three-dimensional
density distribution $\rho(\vec r)$ assuming a dilute, non-interacting ion
cloud at a constant temperature, using $\rho(\vec r) \propto
\exp(-\frac{V(\vec r)}{k_{\rm B} T})$, where $V(\vec r)$ is the trapping
potential as a function of the position $\vec r$, $k_{\rm B}$ is the Boltzmann
constant and $T$ is the absolute temperature. The trapping potential is
assumed to be a superposition of the rf-induced effective potential,
proportional to $r^{20}$, and the static potential of the two end-cap
electrodes \cite{simion}. After integration over the axial coordinate, this
calculation yields the column density (solid line in Fig.\
\ref{particledistr:fig}). In the calculation the effective radius of the
rf-induced potential has been decreased by 10\,\% to match the experimental
width of the ion density, which is assumed to account for deviations from the
assumed symmetry of the trapping field. The calculation reproduces the maxima
of the density qualitatively correct. It also allows to inspect their origin,
which stems from a slight decrease of the column length for small $r$ due to
the repulsive end-cap potential accompanied by a small increase of the
trapping potential of 2-3\,mV. If the $z$-dependence of the potential is
neglected, a rectangular distribution of similar radial extension without the
peaks is obtained, which is shown in Fig.\ \ref{particledistr:fig} as dashed
line. The measured left/right asymmetry shows a deviation from cylindrical
symmetry, which is most likely due to field inhomogeneities introduced by an
asymmetric placement of ground electrodes outside of the trap.

Integration of $k_{\rm pd}(r)$ over the entire trap and division by the
traversing photon flux gives the absolute photodetachment cross section, as
shown in Eq.\ \ref{cross_section:eq}. The integral is evaluated for $r$ to run
from 0 to large radial distances in Fig.\ \ref{particledistr:fig}. Assuming to
first order that cylindrical symmetry holds, despite the observed up-down
asymmetry, the integrals for the left and right half of the distribution are
averaged to derive the cross section. Their difference yields the dominant
contribution to the accuracy of the cross section. We thus obtain for
photodetachment of OH$^-$ at 632.8\,nm an absolute cross section of $(5.6\pm
1.4)\times10^{-18}\,$cm$^2$.

The two previous measurements of the absolute OH$^-$ photodetachment cross
section state values near 630\,nm of 12.5$\times$10$^{-18}\,$cm$^2$
\cite{branscomb1966:pra} and 7.5$\times$10$^{-18}\,$cm$^2$
\cite{lee1979:jcp}. Theoretical work, neglecting internal degrees of freedom,
indicates a cross section smaller than 10$^{-18}$\,cm$^2$ at this wavelength
\cite{soerensen1995:ps}. These values lie above and below of our measured
value and outside of its accuracy.  These deviations can be explained by the
effect of rotational excitation of the OH$^-$ anions. By summing up the
H{\"o}nl-London factors for photodetachment into all accessible final states
\cite{schulz1983:pra,goldfarb2004:jcp}, it turns out that the rotational
contribution to the photodetachment cross section is given by a factor
proportional to $(2J+1)$, where $J$ is the rotational state of OH$^-$. This
leads to a photodetachment cross section ratio for thermal distributions at
0\,K, 170\,K, 300\,K and 850\,K of 1:4.3:5.8:9.8. These ratios explain the
much larger measured values compared to the calculation and they match the
factors between the three measurements, considering that the work of Lee and
Smith was carried out at 300\,K and assuming the unknown rotational
temperature in the experiment of Branscomb amounts to 850\,K, which is
reasonable for the employed discharge ion source.  The notable implication of
this analysis is that internally cold OH$^-$ ions ($J=0$) should have a much
smaller photodetachment cross section which would make them significantly more
stable in interstellar environments.

In conclusion, we have developed and employed a sensitive scheme to measure
absolute photodetachment cross sections of negative ions in a multipole ion
trap. Applied to OH$^-$ anions an absolute cross section is obtained and a
significant dependence on the rotational quantum state of OH$^-$ is
deduced. Furthermore, the column density of the ions in the trap is measured
in situ and is compared to calculations of the trapping potential.

The rotational state dependence of the photodetachment cross section may be
relevant for models of negative ion abundances in interstellar space where low
rotational temperatures prevail. To shed more light on the possible existence
of anions in space we intend to extend our work towards photodetachment at
lower temperatures and of larger molecules such as polyaromatic
hydrocarbons. These experiments will be aided by the high sensitivity of the
presented scheme, which allows to access photodetachment cross sections down
to 10$^{-21}$\,cm$^2$.

\begin{acknowledgments}
This work is supported by the Deutsche Forschungsgemeinschaft and the
Elitef{\"o}rderprogramm der Landesstiftung Baden-W{\"u}rttemberg.
\end{acknowledgments}


\end{document}